# Time-bin Phase and Polarization based QKD systems performance analysis over 16Km Aerial Fibers

Persefoni Konteli *[1], Nikolaos Makris [1], Konstantinos Tsimvrakidis [1], Alkinoos Papageorgopoulos [1], Ilias Papastamatiou [2], Petros Papapetropoulos [2], Dimitrios Syvridis [1], George T. Kanellos [1]

[1] National and Kapodistrian University of Athens, Athens, Greece;
[2] GRNET S.A. – National Infrastructures for Research and Technology, Athens, Greece
perskonteli@di.uoa.gr

***Abstract:*** We analyze the performance of Time-bin Phase and Polarization based QKD systems on mixed 14Km underground and 16Km of aerial fiber using plug-and-play commercial QKD systems.

I. INTRODUCTION

Several Quantum Key Distribution (QKD) protocols have been proposed to secure communications by allowing two remote parties to share a symmetric key using various qubit encoding methods, such as polarization, phase, time-bin or combinations thereof. While optical fibers are the most common medium for transmitting quantum states, they are typically deployed buried underground. Recently aerial fiber installations have been introduced in suburban and rural areas to reduce deployment costs and simplify installation complexity [1]. However, despite the installation advantages, aerial fiber deployments are more susceptible to Polarization Mode Dispersion (PMD), a significant challenge in optical communications. In classical high-bit rate communication systems, PMD causes optical pulse broadening and directly impacts the State of Polarization (SOP). In QKD systems, fluctuations in SOP lead to unpredictable polarization basis misalignments [2], which in turn increases the Quantum Bit Error Rate (QBER). The impact of PMD is more pronounced in aerial fibers due to environmental factors. Wind-induced movements and rapid temperature changes cause dynamic SOP fluctuations along different segments of the aerial fiber line [3],[4]. In contrast, buried and submarine fibers experience minimal SOP variations due to their stable environments that result in slower, more gradual turbulence or temperature changes. QKD systems can compensate for slow SOP changes in real-time, but rapid fluctuations are particularly challenging, especially for QKD protocols that rely solely on polarization-encoded photon states. These dynamic SOP changes can disrupt quantum communication between the authorized parties, leading to increased QBER. To address this, several works have proposed and developed compensation techniques, such as fast polarization feedback modules with sub-millisecond response times [5], multiplexing methods for reference pulses [6] and algorithms for reduced compensation response time [7]. In [8], a plug-and-play QKD system with polarization auto-compensation was used to test the stability of quantum key exchange over 67 km, including 5 km of aerial fiber. Similarly, [5] presented a fast polarization feedback system that enabled quantum communication between two remote parties over 68 km with 15 dB of attenuation, reaching an SKR of 2 kbps. A subsequent study [2] conducted an analysis on polarization fluctuations observed in the same aerial fiber field trial. Additionally, [9] reported on a stable 70-day QKD operation over 61 km of aerial fiber with a total loss of 23 dB, achieving a mean SKR of 1 kbps.

In this work, we report on the impact of aerial optical fiber on QKD systems performance by applying two QKD systems with different encoding methods: one based on polarization and the other on both phase and time-bin encoding. The two commercial QKD systems employed were tested over a 30 km link (9.8 dB of attenuation), consisting of 16 km of aerial fiber and 14 km of subterranean fiber, achieving SKR up to 6 kbps. The QKD system using the phase and time-bin (TBP) encoding, with a power budget of 30 dB, was tested at three different attenuation levels (10, 23 and 30 dB), while the polarization-based (PB) QKD system, with a nominal operational limit of 20 dB, was only tested at 15 dB of line attenuation. The TBP QKD system showed greater stability under varying conditions, which is why it was tested across multiple attenuation levels to assess its performance under stress. On the other hand, the polarization-based system also managed to achieve stable performance, requiring however more frequent initialization process of the protocol. Each configuration was tested for 40 hours, and the results indicate that the TBP system encoding system was able to produce keys for longer time periods than the PB system, highlighting the increased sensitivity of polarization encoding to environmental fluctuations in aerial fiber. Overall terrestrial fibers demonstrate greater stability than the corresponding aerial fiber setups in every encoding method, due to unpredictable SOP changes.

II. EXPERIMENTAL PROCEDURE

The experimental testbed featured two commercially available QKD systems, both implementing the BB84 protocol, ThinkQuantum QUKY with polarization-based encoding [10] and quantum channel operating at 1549.32 nm wavelength

(ITU Channel 35), and IDQuantique Clavis XG with time-bin and phase encoding at its X and Y base respectively [11], operating at 1551.72 nm wavelength (ITU Channel 32). The QKD transmitter (Alice) of both pairs were located at the DLR Data Center (DC) in Koropi, Athens, Greece, while the receiver (Bob) were located at the DLR Data Center in Agia Paraskevi, Athens, Greece, connected via a dedicated fiber pair. Each fiber link is composed of the following sections: a 5.5 km G-657 A1 terrestrial optical fiber, with an attenuation of 1.35 dB; a 16 km segment of aerial optical fiber, with attenuation of 3.25 dB; and an 8 km section of G-652D terrestrial optical fiber, exhibiting an attenuation of 1.99 dB, as shown in Fig. 1. The TBP pair necessitates two separate optical fiber channels for its classical communication needs, both operating bidirectionally at the 1553.33 nm wavelength (ITU Channel 30). Since there was only one extra optical fiber link (apart from the quantum fiber channel), both the TBP system service channels were co-propagating through a single bidirectional link established with the help of circulators, one at each DC. On the other hand, the PB system implements a polarization 3-State efficient BB84 decoy state protocol by utilizing iPOGNAC, a novel polarization encoder that does not require any calibration at the transmitter. iPOGNAC is combined with Qubit Synchronization [10] that allows classical communication to operate via classical network communication and thus requires only one optical fiber link for the transmission of quantum information. The TBP system pair was tested at three attenuation levels, 10 dB, 23 dB and 30 dB and was let to operate under each value for over 40 hours, before switching to the PB system. The total losses of the optical fiber links were measured at 9.8 dB (including the optical fibers, the fiber connectors and the patch cords from and to the DC patch panels), therefore additional optical attenuators were employed to achieve higher attenuation levels for the quantum channel of each QKD system. The PB pair can operate in two modes, depending on the quantum link losses, where when operating in a highly lossy fiber channel under-biases the Bob's single photon detector to decrease the dark count rate and achieve a higher SNR, and therefore an increased Secret Key Rate (SKR). The PB system was operable only at 15 dB of attenuation set in a high loss mode. To assess the impact of the aerial optical fiber on the performance of both QKD systems, we replicated the field deployed conditions in terms of distance and attenuation within our laboratory environment. This was carried out by utilizing a spool of single mode fiber (SMF) that precisely matched the 30 km length used in the field configuration. The total attenuation of the fiber was 8 dB. To accurately simulate the signal loss characteristics, we added appropriate attenuators for each scenario to ensure that the total attenuation matched the real-world deployment conditions, both for the three attenuation levels of the TBP system and the 15 dB case for the PB system. For both the TPB and the PB systems, lab measurements were conducted for over 40 hours.

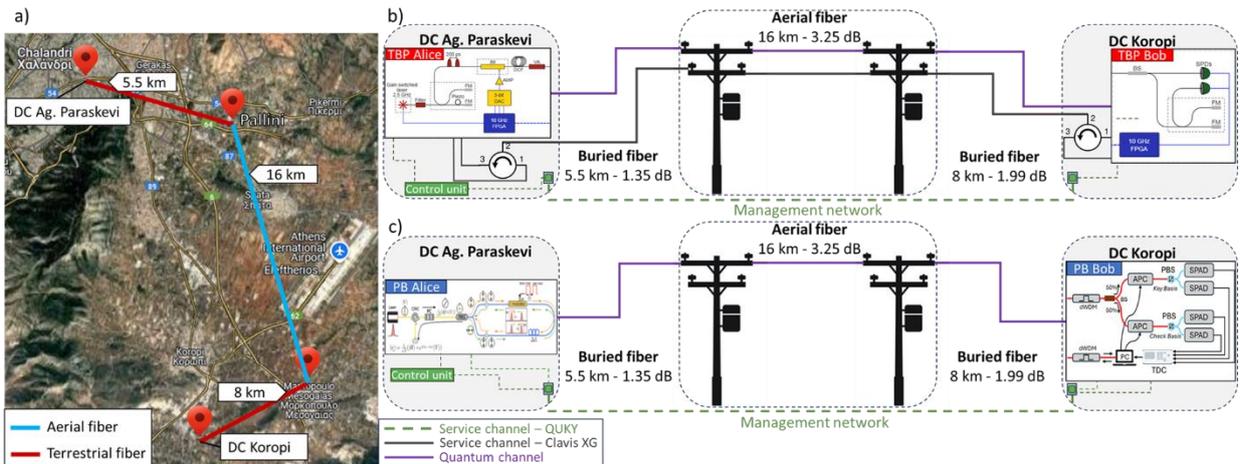

Fig. 1. a) Physical location of the two Data Centers in Athens connected by the aerial and the subterranean fiber segments. Optical and ethernet connections of the experimental setup for b) TBP system and c) PB system.

## III. RESULTS

We report on the performance of the TBP and PB systems respectively by monitoring their SKR (Fig. 2) and QBER (Fig. 3) in various attenuation levels both for the in-lab scenario and the field-trial transmission. However, for multiple attenuation levels, the PB system was unable to pass the Base agreement phase (initialization procedure of the bases X and Z) in a reasonable amount of time because of the system exhibiting high basis misalignment due to constant fluctuations of SOP so we restricted our study to 15 dB for the PB system. In the aerial fiber setup, initialization required significantly more time compared to the in-lab fiber setup due to continuous changes in SOP which affected the base agreement process to properly align, requiring several iterations to reach the 16 dB threshold. For the TBP system the initialization procedure remained consistent at all attenuation levels, lasting approximately 10 minutes in both cases. After initialization, both TBP pair (for all attenuation levels) and the PB pair were operating for over 40 hours. The SKR performance for each scenario is shown in Fig. 2a-d. During operation, the quantum key generation for the PB-system over the aerial fiber was disrupted 35 times accounting for 24.82% of the total operation time due to SOP changes. In contrast, the in-lab fiber setup exhibited perfect stability without any disruptions. The 0 bps SKR drops observed at all 3 attenuation levels for TBP system in the aerial setup (Fig. 2b), are attributed to service channel loss (correlated from logs), a behavior attributed to the instability of the service channel over the aerial fiber potentially caused by reflections both from the circulators and the fiber patch cords used. As expected, for the TBP pair performance degrades with higher attenuation and stable operation is observed in every tested attenuation level, both for the aerial and in-lab fiber setups. On the other hand, the PB pair successfully produced keys in the in-lab setup, but in the mixed fiber with aerial segment setup the operation was disrupted for 10.8 hours. In this period the system spent most of the time in base agreement stage attempting to reinitialize the QBER bases while secure key generation ceased, producing zero values as shown in Fig. 2d.

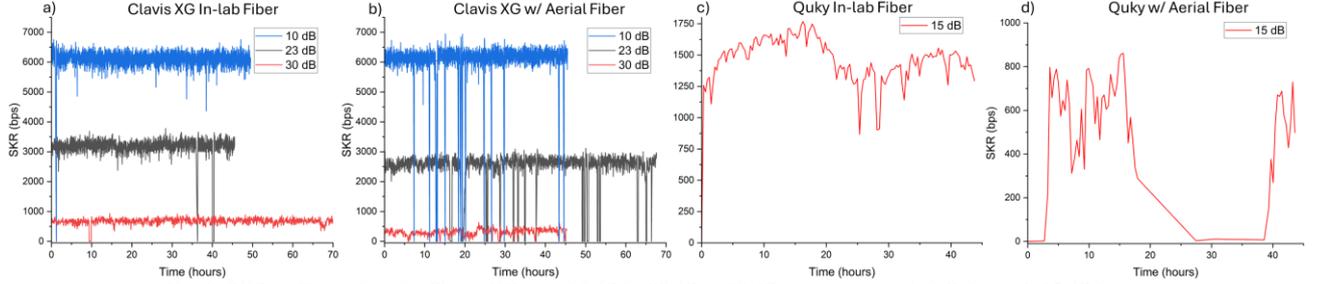

Fig. 2. SKR vs Time plots for Clavis XG in a) 10 dB b) 23 dB c) 30 dB attenuation and d) Quky with 15 dB loss.

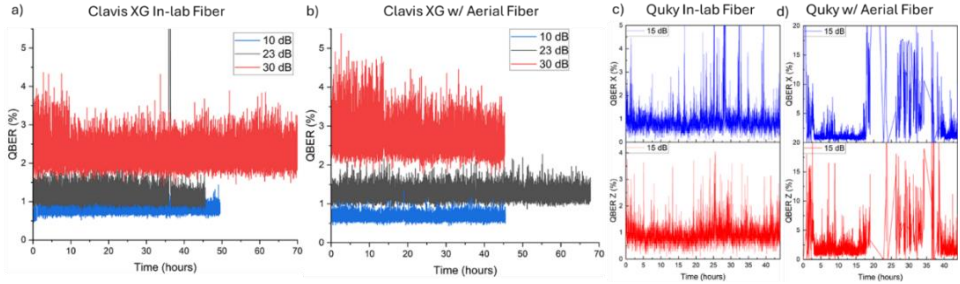

Fig. 3. QBER vs Time plots for Clavis XG in a) 10 dB b) 23 dB c) 30 dB attenuation and d) Quky with 15 dB loss.

TABLE I. Mean values comparison between aerial and in-lab fiber setups for each metric. The extent at which the aerial fiber affects communication for each device can be identified for each scenario.

| System | Attenuation | SKR-In-lab Fiber (bps) | SKR-w/ Aerial Fiber (bps) | QBER-In-lab Fiber (%) | QBER-w/ Aerial Fiber (%) |
|---|---|---|---|---|---|
| TBP | 10 dB | 6114.90 | 6114.60 | 0.80 | 0.66 |
| TBP | 23 dB | 3190.34 | 2536.05 | 1.33 | 1.26 |
| TBP | 30 dB | 675.75 | 312.99 | 2.22 | 2.70 |
| PB | 15 dB | 1458.23 | 538.44 | X: 1.29 Z: 1.08 | X: 2.76 Z: 2.69 |

To make a direct comparison between the mixed fiber setup and in-lab setup TABLE I presents the mean values for SKR and QBER metrics for all tested scenarios. Evidently the TBP system performance was unaffected for the 10 dB low attenuation level, as confirmed by the almost identical mean values of SKR. However, in 23 dB and 30 dB attenuation levels, TBP system performance in the aerial setup degrades in respect to the corresponding in-lab fiber scenario. In fact, for the 30 dB attenuation level, the performance drop corresponds to a 50% drop of SKR. For the PB system, the performance drop during operation phase also drops by 50%, but as the system performance was disrupted for 24.82% of the testing time it indicates greater sensitivity to the aerial setup conditions. This is further confirmed by increase of QBER in both bases as shown in Fig. 3c,d. Overall, the increased stability of the TBP system can be attributed to the employed protocol: The TBP system utilizes a combination of time-bin and phase encoding, where one base uses time-bin encoding while the other encodes the relative phase between two consecutive pulses [11]. Even though polarization changes can affect the time-bin base because of changes in SOP, the other base in not affected since the same effect will occur in both consecutive pulses. On the other hand, the PB system encoder relies only on polarization encoding. This could result in increased vulnerability to steep polarization changes, requiring a fast polarization compensation to correct basis misalignment.

## IV. CONCLUSIONS

We report on a successful demonstration of QKD distribution over mixed optical fiber links that included 16 km of aerial optical fiber segments, using two commercial QKD pairs with different qubit encoding schemes. The phase and time bin-based QKD system was evaluated at 3 attenuation levels (10, 23, and 30 dB) showing stable operation across all levels revealing that for operation in higher attenuation levels the SKR and QBER metrics were affected over the aerial fiber when compared to in-lab fiber spool scenarios. The polarization-based QKD system required more frequent initialization compared to an equivalent in-lab setup without aerial segments. Also, the PB system operation managed to demonstrate successful key generation, but the process was disrupted for 24.82 % of the test duration, showing greater sensitivity to the changes in SOP.

## ACKNOWLEDGMENTS

This work was funded by the EC HellasQCI (GA 101091504), LaiQa (GA 101135245), and QRONOS (GA 16826).